%% file: Main.tex
\documentclass[11pt,english]{article}
\usepackage{import}
\usepackage[utf8]{inputenc}
\usepackage[T1]{fontenc}
\usepackage[a4paper,       
            bindingoffset=0.2in,
            left=0.50in,
            right=0.50in,
            top=0.85in,
            bottom=0.85in,
            footskip=.25in]{geometry}
\usepackage{amsmath}
\usepackage{amsthm}    
\usepackage{amssymb}  
\usepackage{bm}       
\usepackage{moreverb}
\usepackage{graphicx} 
\graphicspath{{Files/Figures/}} 
\usepackage{accents}
\usepackage{epstopdf}
\usepackage{float}
\usepackage{footnote}
\usepackage{threeparttable}
\usepackage{breakcites}
\usepackage{blindtext}
\usepackage[parfill]{parskip}
\usepackage{multirow}
\usepackage{booktabs}
\usepackage{hyperref}
\usepackage{authblk}
\usepackage{xcolor}
\usepackage[toc,page,title]{appendix} 
\usepackage[backend=bibtex,style=numeric, sorting=none]{biblatex} 
\usepackage{breqn}
\usepackage{titlesec}
\usepackage{comment} 
\usepackage{subfiles} 
\titleformat{\paragraph}
{\normalfont\normalsize\bfseries}{\theparagraph}{1em}{}
\titlespacing*{\paragraph}
{0pt}{3.25ex plus 1ex minus .2ex}{1.5ex plus .2ex}
\DeclareUnicodeCharacter{2060}{\nolinebreak}
\makesavenoteenv{tabular}
\begin{document}


\subfile{Files/Abstract.tex}

\newpage

\subfile{Files/Manuscript.tex}

\newpage

\subfile{Files/Highlights_acknl_data.tex}

\newpage
\printbibliography

\setcounter{figure}{0} 
\clearpage 
\clearpage

\begin{appendices}

\renewcommand{\theequation}{A.\arabic{equation}}
\numberwithin{equation}{section}
\setcounter{equation}{0}  

\subfile{Files/Appendix.tex}

\end{appendices}

\end{document}

%% file: Files/Abstract.tex
\title{\LARGE{ MetaBayesDTA: Codeless Bayesian meta-analysis of test accuracy, with or without a gold standard }}

\author[1,2]{ Enzo Cerullo }
\author[1,2]{ Alex J. Sutton }
\author[3]{ Hayley E. Jones }
\author[2]{ Olivia Wu }
\author[2,4]{Terry J. Quinn }
\author[1,2]{ Nicola J. Cooper }

\affil[1]{\small{Biostatistics Research Group, Department of Population Health Sciences, University of Leicester, Leicester, UK}}
\affil[2]{Complex Reviews Support Unit, University of Leicester \& University of Glasgow, Glasgow, UK}
\affil[3]{Population Health Sciences, Bristol Medical School, University of Bristol, Bristol, UK}
\affil[4]{Institute of Cardiovascular and Medical Sciences, University of Glasgow, Glasgow, UK }

\date{}

\maketitle


\section*{\large{Abstract}}

\textbf{Introduction}: Despite their applicability, statistical models used for the meta-analysis of test accuracy require specialised knowledge to implement, with the necessary level of expertise having recently increased. This is due to the development and recommendation to use more sophisticated methods; such as those in Version 2 of the Cochrane Handbook for Systematic Reviews of Diagnostic Test Accuracy. This paper describes a web-based application that extends the functionality of previous applications, making many advanced analysis methods more accessible.  

\textbf{Methods}: We sought to create an extended, stand-alone, Bayesian version of MetaDTA, which (i) has the benefits of previously proposed applications and addresses key limitations of them, (ii) is accessible to researchers who do not have the specific expertise required to fit such models, and (iii) is suitable for experienced analysts. We created the application using Shiny and Stan. 

\textbf{Results}: We created MetaBayesDTA (\url{https://crsu.shinyapps.io/MetaBayesDTA/}), which allows users to conduct meta-analysis of test accuracy, with or without a gold standard. The application addresses several key limitations of other applications. For instance, for the bivariate model, one can conduct subgroup analysis, univariate meta-regression, and comparative test accuracy evaluation. Meanwhile, for the model which does not assume a perfect gold standard, the application can account for the fact that studies use different reference tests.

\textbf{Conclusions}: Due to its user-friendliness and broad array of features, MetaBayesDTA should appeal to a wide variety of researchers. We anticipate that the application will encourage wider use of more advanced methods, which ultimately should improve the quality of test accuracy reviews.

\section*{\large{Keywords}}

Meta-Analysis, diagnostic test accuracy, application, imperfect gold standard, latent class

*Corresponding Author

Email address: enzo.cerullo@bath.edu

%% file: Files/Manuscript.tex
\section{Introduction}
\label{section_introduction}

\subsection{Background to meta-analysis of test accuracy}
\label{section_background}
In medicine, tests are used to screen, monitor and diagnose medical conditions, and therefore it is imperative that these tests produce accurate results. This 'accuracy' refers to their sensitivity and specificity. The former is the probability that a test can correctly identify patients who have the disease and the latter is the probability that the test can correctly identify patients who do not have the disease. To evaluate their accuracy, studies and analyses are carried out to compare the results of the test under evaluation (called the 'index' test) against some existing test, which is assumed to be perfect (called the 'reference' or 'gold standard' test). Index tests typically have lower accuracy than the gold standard; however, they are often quicker, cheaper and/or less invasive. 

Standard methods for the meta-analysis of test accuracy assume that the gold standard test is perfect - i.e., that the test is 100\% sensitive and specific. These models dichotomize the data into diseased and non-diseased according to the results of the reference test, and include the bivariate model of Reitsma et al \supercite{Reitsma2005} and the hierarchical summary receiver operating characteristic (HSROC) model of Rutter \& Gatsonis \supercite{Rutter2001}. These models have been shown to be equivalent in practice when no covariates are included \supercite{Harbord2007}. 
Models which do not assume a perfect gold standard have also been developed \supercite{Chu2009, Menten2013, dendukuri2012}. These models - which are often referred to as \textit{traditional latent class models} (TLCMs) - assume that each test is measuring the same latent disease, and each individual is assumed to belong in either the diseased or non-diseased classes. These methods can also model the correlation between each test within each disease class (i.e. the \textit{conditional dependence} between tests). All of the aforementioned methods take into account the correlation between sensitivity and specificity across studies.

\subsection{Why is this application needed?}
\label{section_why_app_needed}
The models discussed in section \ref{section_background} require statistical programming expertise using software such as R or Stata. Cochrane, an organisation who help support evidence-based decisions about health interventions such as diagnostic and screening tests. Whilst they do provide free software  Revman \supercite{cochrane_revman} using the Moses-Littenberg method \supercite{littenberg_moses_1993}, it fails to appropriately account for random effects and the correlation between sensitivity and specificity across studies. Carrying out meta-analysis of test accuracy using online applications has a lower user burden since no programming is needed. Not only does this make such methods accessible to a broader array of people, it also streamlines the workflow for more experienced data analysts. 

Other web applications for the meta-analysis of test accuracy include MetaDTA \supercite{metadta_Freeman2019-mk} (\url{https://crsu.shinyapps.io/dta_ma/}), which provided the foundation for our new app, and BayesDTA (\url{https://bayesdta.shinyapps.io/meta-analysis}). The former uses frequentist methods and implements the bivariate model \supercite{Reitsma2005}, allowing for risk of bias and quality assessment data to be incorporated into the results plots. The latter uses Bayesian methods and incorporates both the bivariate model \supercite{Reitsma2005} and the TLCM model \supercite{Chu2009, Menten2013}. Similarly to BayesDTA, our application, MetaBayesDTA (\url{https://crsu.shinyapps.io/MetaBayesDTA/}), runs Bayesian versions of both the bivariate \supercite{Reitsma2005} and the TLCM model \supercite{Chu2009, Menten2013}, and is powered by Stan \supercite{stan}, a Bayesian model fitting software. However, unlike BayesDTA, our application can also conduct subgroup analysis and meta-regression for the bivariate model, and can be used to conduct a comparative meta-analysis of test accuracy for 2 or more tests using categorical meta-regression (assuming the same variances between tests), using methods recommended in chapter 11 of version 2 of the Cochrane handbook for systematic reviews of diagnostic test accuracy\supercite{cochrane_dta_handbook_v2}. Furthermore, for the TLCM model, rather than assuming all studies use the same reference tests, it can model multiple reference tests.  A full comparison between MetaBayesDTA, MetaDTA and BayesDTA is shown in table \ref{table_app_comparison}.

\begin{table}[H]
\caption{Table comparing features of MetaBayesDTA, MetaDTA and BayesDTA}
\begin{tabular}{|l|l|l|l|}
\hline
                                                                  & MetaBayesDTA              & MetaDTA                   & BayesDTA                  \\ \hline
Bayesian or frequentist                                           & Bayesian                  & Frequentist               & Bayesian                  \\ \hline
Model assuming gold standard (bivariate model)                    & $\checkmark$ & $\checkmark$ & $\checkmark$ \\ \hline
For bivariate - subgroup analysis                                 & $\checkmark$ & $\times$     & $\times$     \\ \hline
For bivariate - univariate meta-regression                        & $\checkmark$ & $\times$     & $\times$     \\ \hline
For bivariate - comparative accuracy of 2+ tests                  & $\checkmark$  & $\times$     & $\times$     \\ \hline
Model not assuming perfect gold standard (TLCM)                   & $\checkmark$ & $\times$     & $\checkmark$     \\ \hline
For TLCM - model multiple reference tests                         & $\checkmark$ & $\times$     & $\times$     \\ \hline
For TLCM - subgroup analysis                                      & $\times$     & $\times$     & $\times$     \\ \hline
For TLCM - univariate meta-regression                             & $\times$     & $\times$     & $\times$     \\ \hline
For TLCM - model fit (correlation residual plot)                  & $\checkmark$ & $\times$     & $\times$     \\ \hline
Interactive layout / pop-up menus                                 & $\checkmark$ & $\times$     & $\times$     \\ \hline
Risk of bias and quality assessment on plots                      & $\checkmark$ & $\checkmark$ & $\times$     \\ \hline
Appropriate restrictions in place                                 & $\checkmark$ & $\times$     & $\times$     \\ \hline
\label{table_app_comparison}
\end{tabular}
\end{table}

\section{Aims}
\label{section_aims}
Our objective was to make an updated, Bayesian version of the R shiny app MetaDTA \supercite{Patel_RSM_graphics_sROC_MA_DTA, shiny_r_package}, which would be accessible to researchers and enable them to conduct a (relatively) robust Bayesian statistical analysis for meta-analysis of test accuracy, despite not possessing sufficient experience in R \supercite{R_software_ref} and Stan \supercite{stan}. It is also aimed at researchers who can use R and Stan (e.g. some data analysts, statisticians, clinical researchers, etc) but would still want to use a web application for efficiency.

\section{Software \& implementation}
\label{section_software_and_implementation}
We used the statistical programming language R \supercite{R_software_ref} to create our web application, using a variety of packages. One such package includes Shiny \supercite{shiny_r_package}, which enables R users to create web applications without having to have knowledge of web development languages such as HTML and JavaScript. Another package used includes rstan \supercite{rstan_r_package}, which enables users to fit Bayesian statistical models in R using Stan \supercite{stan}, and is what we used to fit both the bivariate and TLCM models in the application. 

\section{Demonstration using motivating example}
\label{section_app_demonstration_motivating_example}
A new user interface format was developed using the R packages shinydashboard \supercite{shinydashboard_r_package} and shinywidgets \supercite{shinywidgets_r_package}. This allows the app to have a clean layout, with many of the menus hidden unless the user chooses to display them. 

\subsection{Data} 
\label{section_data_tab}
\begin{figure}[H]
    \centering
    \includegraphics[width=15cm]{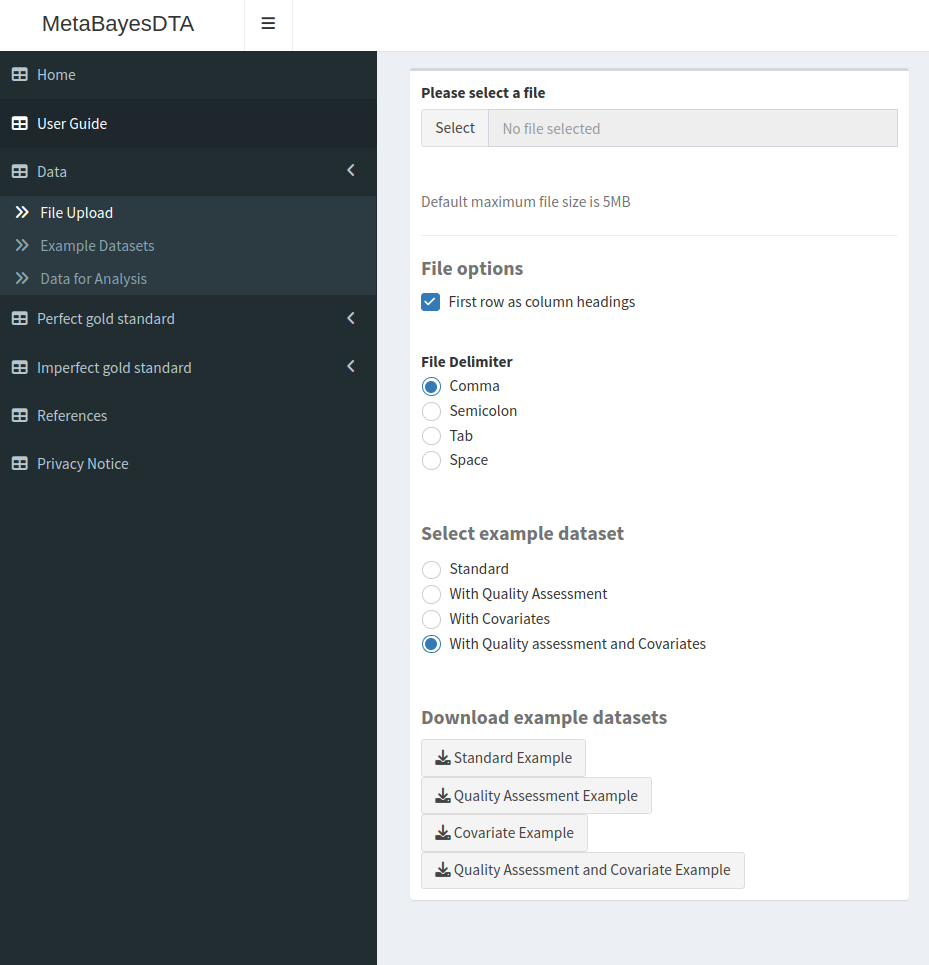}
    \caption{Screenshot of 'Data' tab, showing part of the 'File Upload' subtab.}
\label{figure_data_tab}
\end{figure}

The 'Data' tab (see figure \ref{figure_data_tab}) allows users to upload their data. The number of columns the datasets must have will vary depending on whether quality assessment data and/or covariate data is included. Datasets involving no quality assessment or covariate data will have six columns, those involving quality assessment data thirteen, those involving covariates at least seven, and those involving both quality assessment and covariate data at least fourteen. The quality assessment data which can be included is from quality assessment carried out using the QUADAS-2 tool \supercite{quadas_2_tool}. This tool has four domains: (i) patient selection, (ii) index test, (iii) reference standard and (iv) flow of patients through the study and timing of the index test(s) and reference standard. 

The 'File Upload' subtab is pre-loaded with an example dementia dataset from a Cochrane meta-analysis \supercite{iqcode_cochrane_meta_Harrison15}, which is described in more detail in the 'Example datasets' subtab in the application. The 'Data for Analysis' subtab shows the dataset currently being used. 

We used this dataset to demonstrate the application here. This meta-analysis assessed the accuracy of the Informant Questionnaire on Cognitive Decline in the Elderly (IQCODE). This is a screening test used to detect adults who may have clinical dementia within secondary care settings. Thirteen studies were included in the meta-analysis. 
To analyse the data, the Cochrane meta-analysis \supercite{iqcode_cochrane_meta_Harrison15} used the bivariate model and found a pooled summary estimates of 0.91 (95\% CI = (0.86, 0.94)) and 0.66 (95\% CI = (0.56, 0.75)) for the sensitivity and specificity, respectively.

\subsection{Perfect gold standard}
\label{section_perfect_gold_standard}
The 'Perfect gold standard' page consists of three tabs: meta-analysis, meta-regression and subgroup analysis. All three tabs use the bivariate model proposed by Reitsma et al \supercite{Reitsma2005}, employing the variation which uses binomial likelihoods proposed by Chu and Cole \supercite{chu_cole_2006_bivariate_2}.  

\subsubsection{Meta-analysis}
The Meta-analysis subtab is split into two halves - the left half consists of the following tabs: 'priors', 'run model', 'study-level outcomes', 'parameter estimates', 'parameters for RevMan', and 'model diagnostics'. The right half has the tabs 'sROC plot', 'Forest Plots' and Prevalence'. 

\begin{figure}[H]
    \centering
    \includegraphics[width=15cm]{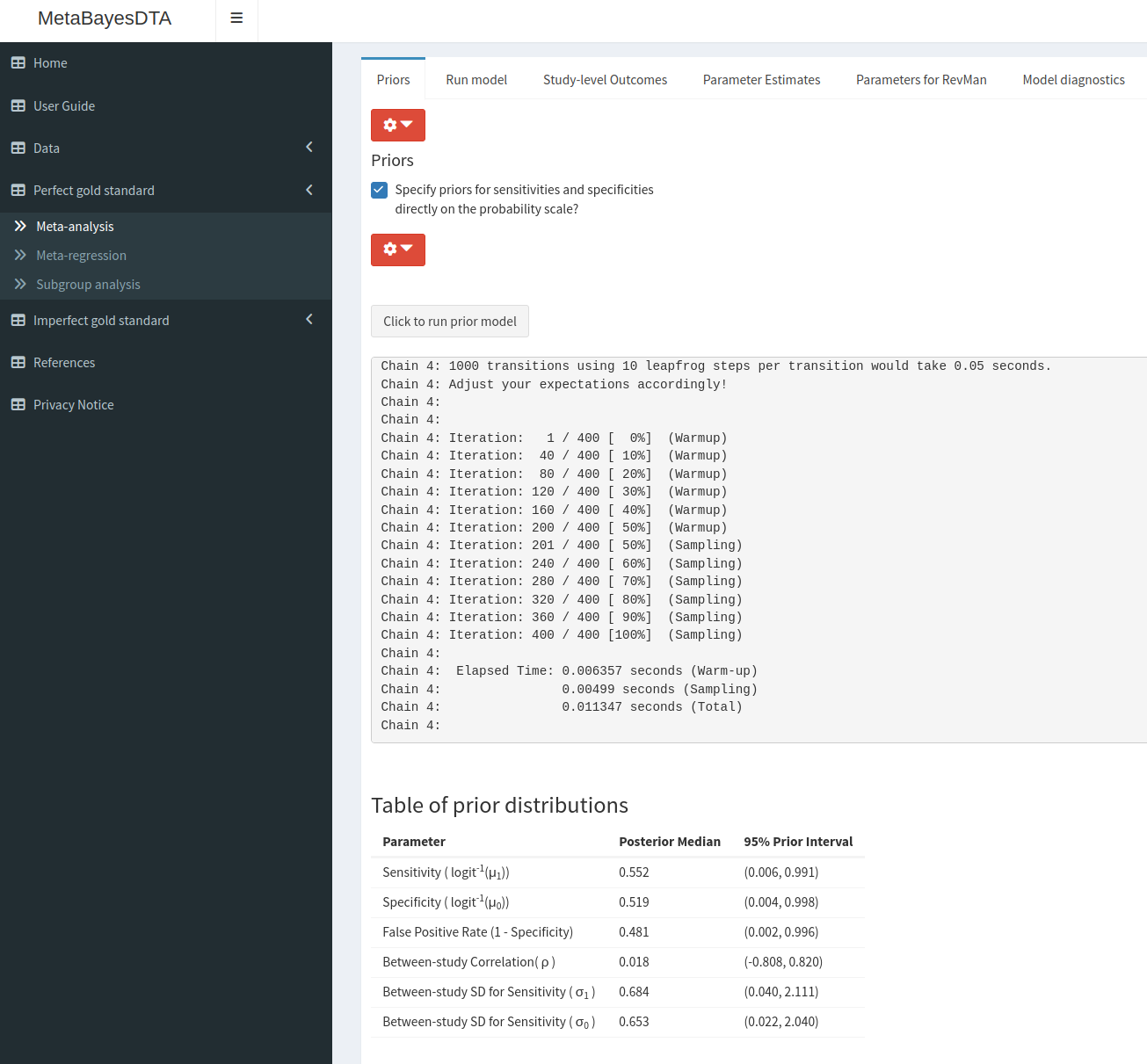}
    \caption{Screenshot of 'Perfect gold standard' tab, 'Priors' subtab within 'Meta-analysis' subtab}
\label{figure_perfect_gs_meta_analysis_priors_segment}
\end{figure}

Since all of the models in the app are Bayesian, prior distributions need to be specified. The 'priors' subtab (see figure \ref{figure_perfect_gs_meta_analysis_priors_segment}) is where users specify prior distributions. The priors can be changed if some information is known about them, and they can be specified in terms of the logistic-transformed ("logit") sensitivity and specificity, or directly on the probability scale. The default prior distributions are weakly informative. More specifically, for the pooled logit sensitivity and logit specificity, we used a normal distribution with mean zero and SD of 1.5 ($N(0, 1.5)$), which is equivalent to a 95\% prior interval of $(0.05,0.95)$ on the probability scale. For the between-study SD's we used a truncated (at zero) normal with zero mean and unit SD ($N_{ \ge 0 }(0, 1.5) $), and for the between-study correlation we used an LKJ \supercite{Lewandowski2009} prior with shape parameter of 2 ($LKJ(2)$), which gives a 95\% prior interval of $(-0.8, 0.8)$. 

Users can examine the prior distributions specified by clicking on the button 'Click to run prior model' and the prior medians and 95\% prior intervals are shown in a table (see bottom of figure \ref{figure_perfect_gs_meta_analysis_priors_segment}). Plots of the prior distributions are also displayed (below the table - not shown in figure \ref{figure_perfect_gs_meta_analysis_priors_segment}). 

Users can run the model by clicking on the 'Click to run model' button within the 'Run model' subtab. In this subtab, users can also run sensitivity analysis - more specifically, this is where any number of studies can be excluded from the analysis to assess the influence of particular studies on the overall pooled estimates. 

\begin{figure}[H]
    \centering
    \includegraphics[width=15cm]{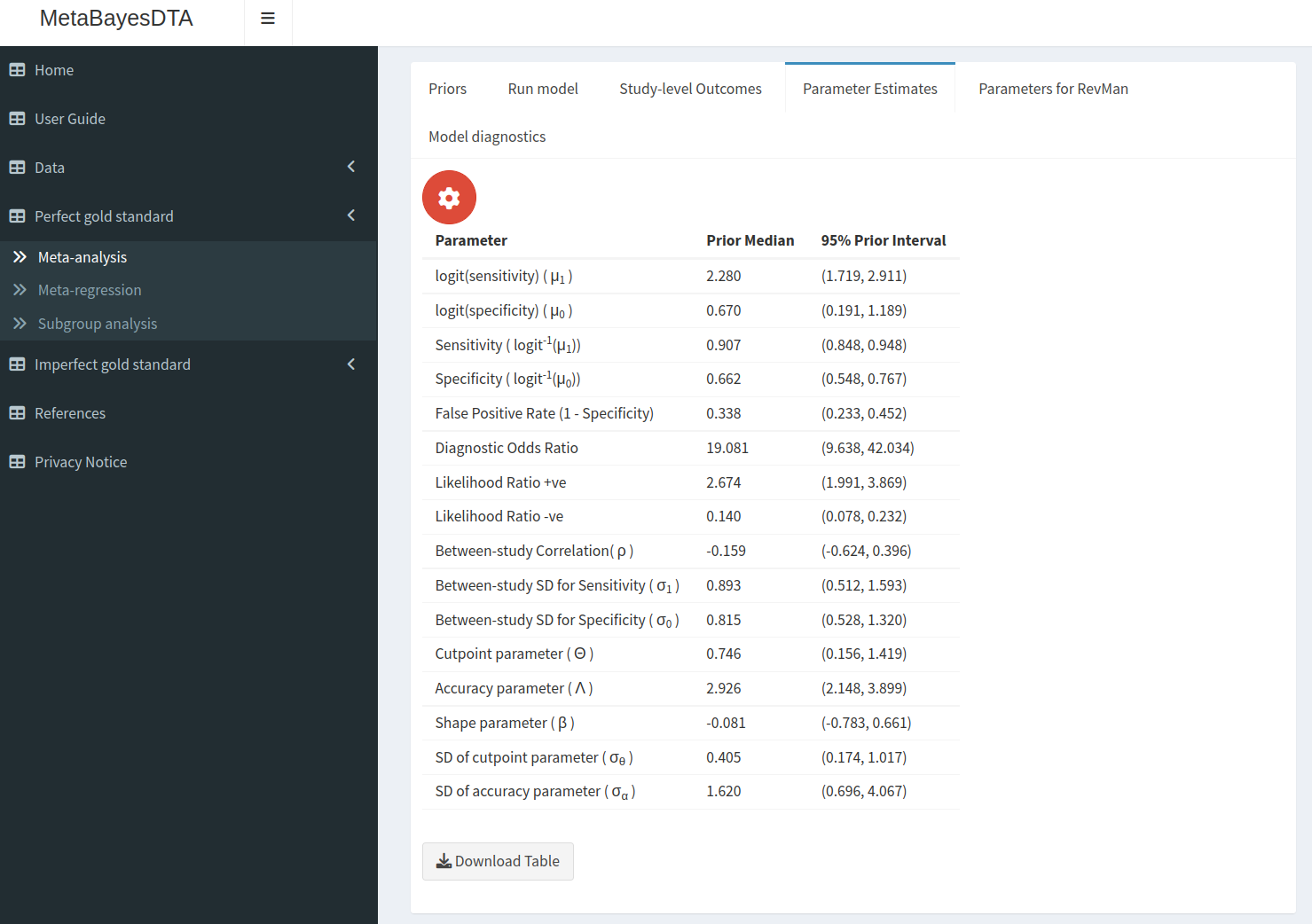}
    \caption{Screenshot of 'Perfect gold standard' tab, 'Parameter estimates' subtab within 'Meta-analysis' subtab}
\label{figure_perfect_gs_meta_analysis_parameter_estimates_segment}
\end{figure}

The 'study-level outcomes' subtab displays key study information that is also displayed in the 'Data' tab, as well as the sensitivity and specificity in each study and study weights - that is, the amount that each study contributes to the overall sensitivity and specificity estimate, calculated using the method from Burke et al \supercite{burke_et_al_study_weights}. 
The 'parameter estimates' subtab (see figure \ref{figure_perfect_gs_meta_analysis_parameter_estimates_segment}) consists of a table with the posterior medians and 95\% posterior intervals for key summary parameters including logit sensitivities and specificities, diagnostic odds ratio and likelihood ratios, between-study correlation and standard deviations, and HSROC parameters. The HSROC parameters are estimated from the bivariate model parameters using the relations shown in Harbord et al \supercite{Harbord2007}. 

The 'parameters for RevMan' subtab consists of the parameter estimates (posterior medians) needed by Cochrane's RevMan software to build ROC plots for people who want to include the analysis results in a Cochrane review. 
The 'Model diagnostics' subtab contains important diagnostics that users must check to ensure whether the model is valid. These include the Stan sampler diagnostics \supercite{Betancourt2017_hmc_intro, stan} - divergent transitions and iterations which have exceeded the maximum treedepth (these should both be 0), split R-hat statistics (should be less than 1.05), and posterior density and trace plots \supercite{stan}. 

\begin{figure}[H]
    \centering
    \includegraphics[width=15cm]{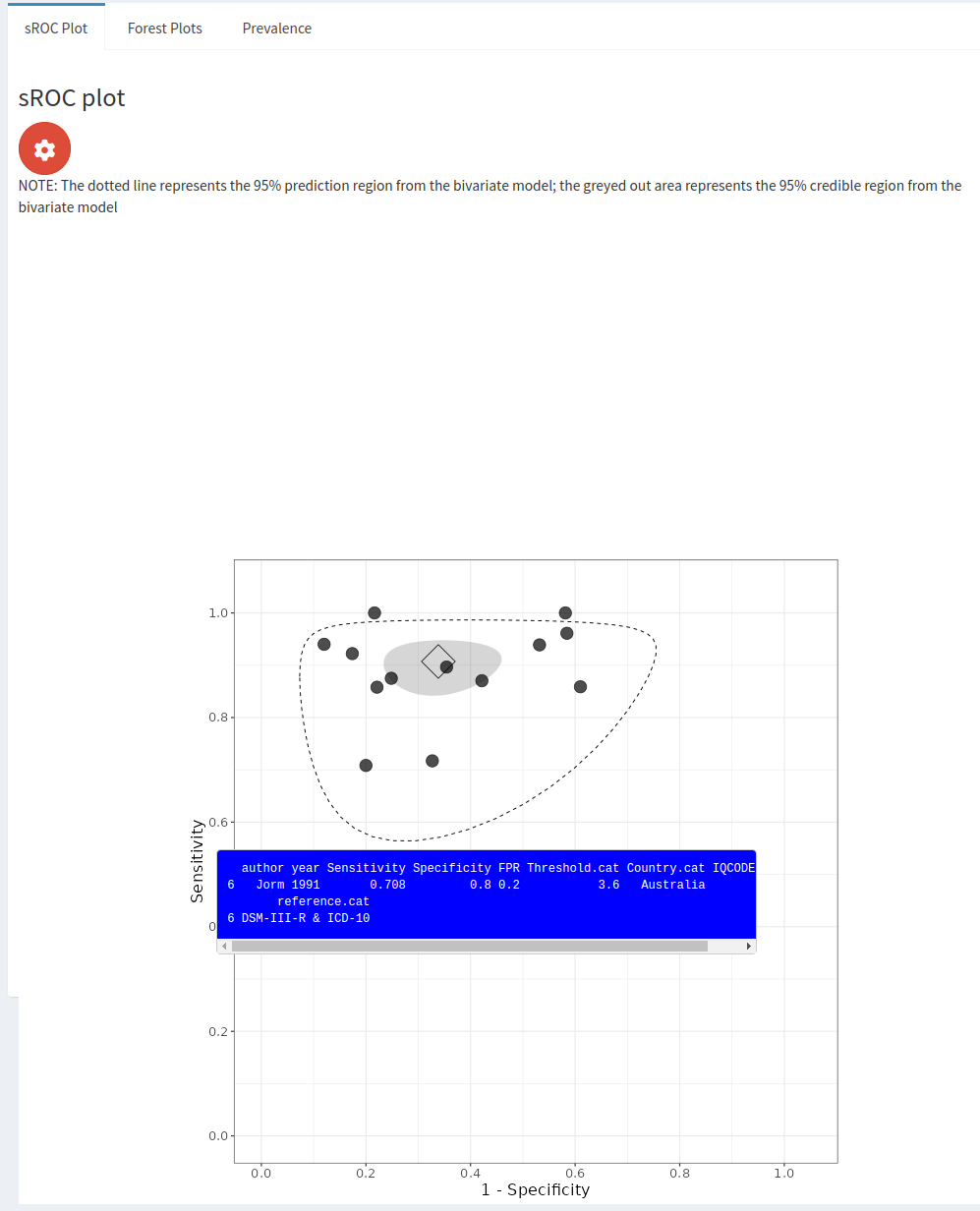}
    \caption{Screenshot of 'Perfect gold standard' tab, 'sROC Plot' subtab within 'Meta-analysis' subtab. The blue box contains information for the study (Jorm et al) corresponding to the point on the bottom-left, and appears when the user clicks on this point.}
\label{figure_perfect_gs_meta_analysis_sroc_plot}
\end{figure}

The sROC plot is displayed in the 'sROC plot' subtab (see figure \ref{figure_perfect_gs_meta_analysis_sroc_plot}). This plot displays the summary estimates, 95\% credible and prediction regions and study-specific sensitivities and specificities. The plot has a range of customization options; for instance, it allows users to change the size of the summary estimates and study-specific points, display the sROC curve, disease prevalence and percentage study weights of each study. It is also interactive - users can click on the study-level points and study-level information will appear over the plot - this is demonstrated in figure \ref{figure_perfect_gs_meta_analysis_sroc_plot}, where the bottom-left point corresponding to the Jorm et al \supercite{jorm_scott_cullen_mackinnon_1991} study has been clicked on. This plot, as well as the other plots produced by the application, can be downloaded. Risk of bias and quality assessment information, if available in the dataset, can also be displayed on the plot (see supplementary material figure \ref{appendix_fig_1}).

The 'forest plots' subtab contains the forest plots - which are plots showing the sensitivity and specificity in each study as well as the corresponding 95\% confidence intervals. 
The 'prevalence' subtab contains a tree diagram which puts the summary estimates into context - it shows how many patients would test positive and negative for a given disease prevalence, and then out of those who test positive and negative, which are diseased and non-diseased. There is also another tree diagram option, which first splits the population by disease status and then by test result. 

We analysed the data discussed in section \ref{section_data_tab} using our application, using the Bayesian bivariate model assuming a perfect gold standard. We used the default prior distributions (see figure \ref{figure_perfect_gs_meta_analysis_priors_segment}), and obtained virtually the same results as the frequentist analysis conducted in the original study - sensitivity and specificity estimates of 0.91 (95\% credible interval [CrI] = (0.85, 0.95)) and 0.66 (95\% CrI = (0.55, 0.77)), respectively (see figure \ref{figure_perfect_gs_meta_analysis_parameter_estimates_segment}). An sROC plot showing the results is shown in figure \ref{figure_perfect_gs_meta_analysis_sroc_plot}.

\subsubsection{Meta-regression}
\label{section_perfect_gs_meta_regression}
The 'Meta-regression' tab is where users can run the bivariate model including a categorical or continuous covariate in an attempt to explain any between study heterogeneity, and consists of subtabs similar to the 'Meta-analysis' tab. The 'Run model', 'study-level outcomes', 'Model Diagnostics' and 'sROC plot' subtabs are the same as those in the 'Meta-analysis' tab. 
Rather than a 'priors' subtab, it has a 'Model set up \& priors' tab, since users also need to select the covariate to use. Furthermore, if using a continuous covariate, users need to specify the value to use for centering (the default is the mean of the values of the covariate) and which value of the covariate to calculate the summary accuracy estimates at. For the default priors, for continuous meta-regression we used $N(0,1.5)$ priors for the pooled logit sensitivity and specificity intercepts and $N(0,1)$ priors for the pooled logit sensitivity and specificity coefficients. For categorical meta-regression, we used $N(0,1)$ priors for the pooled logit sensitivities and specificities for each level of the covariate. For both continuous and categorical meta-regression, similarly to the model with no covariates (see section \ref{section_perfect_gold_standard}), we used ($N_{ \ge 0 }(0, 1.5) $) and $LKJ(2)$ priors for the between-study SD's and correlations, respectively. 

\begin{figure}[H]
    \centering
    \includegraphics[width=15cm]{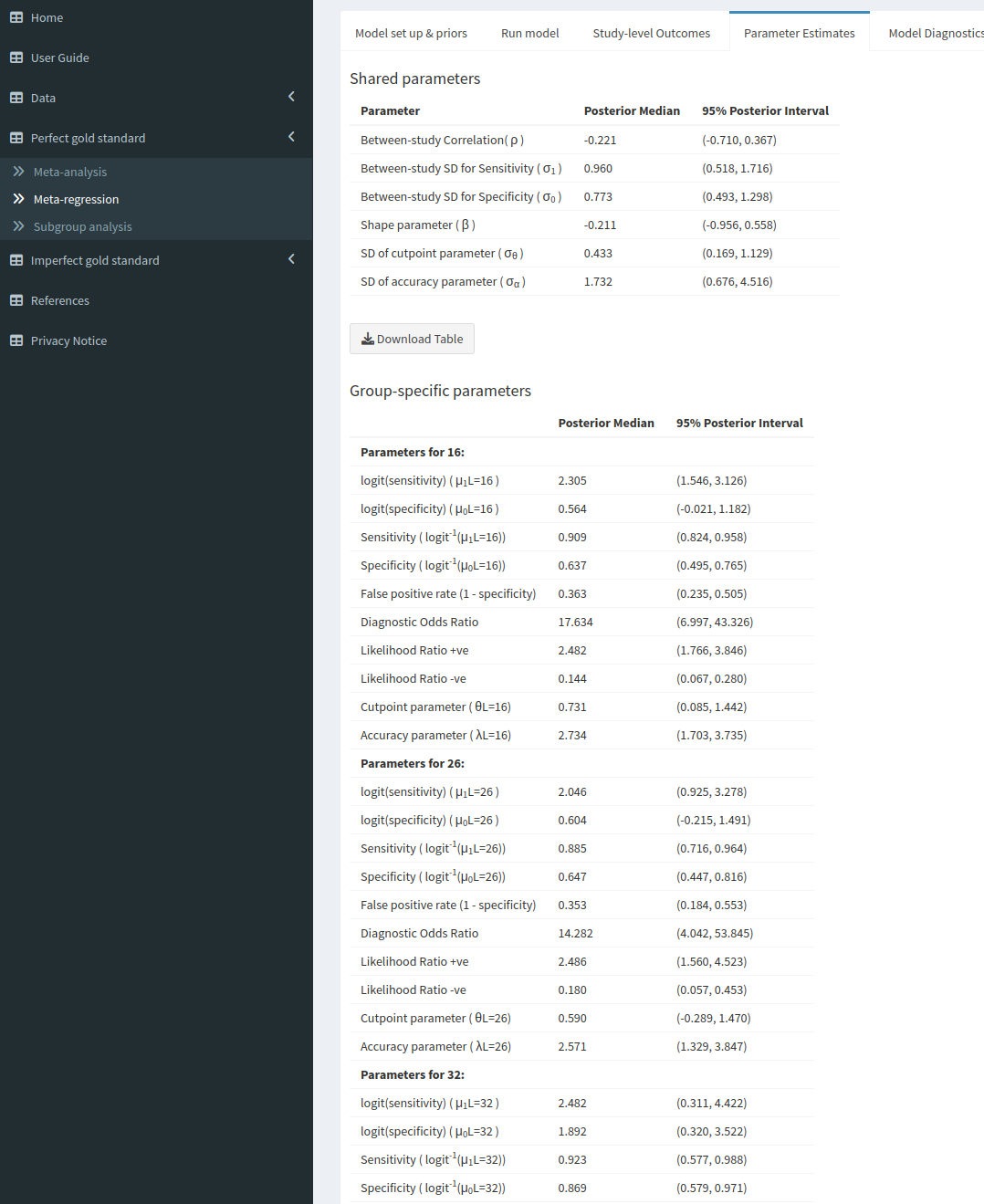}
    \caption{Screenshot of 'Perfect gold standard' tab, 'parameter estimates' subtab within the 'meta-regression' subtab}
\label{figure_perfect_gs_meta_regression_parameter_estimates_segment}
\end{figure}

\begin{figure}[H]
    \centering
    \includegraphics[width=10cm]{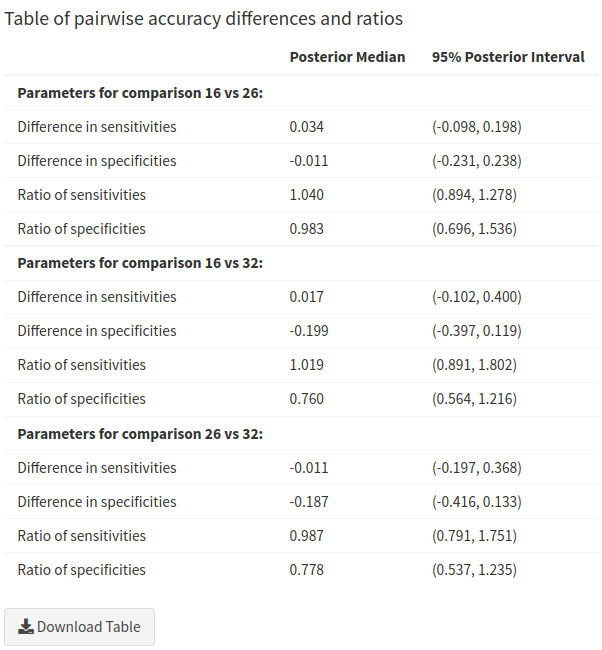}
    \caption{Screenshot of the table of pairwise accuracy differences and ratios table; in the 'parameter estimates' subtab within the 'meta-regression' subtab}
\label{figure_perfect_gs_meta_regression_parameter_estimates_pairwise_diffs_segment}
\end{figure}

The 'parameter estimates' subtab contents will vary depending on whether continuous or categorical meta-regression is being carried out. For the continuous meta-regression, there will be one table showing the parameters which do not vary, regardless of what the user chooses for the covariate value to calculate the summary estimates at, and another table containing the parameters that do vary. For categorical meta-regression (see figure \ref{figure_perfect_gs_meta_regression_parameter_estimates_segment}), there will be one table containing the parameters shared between studies, such as between-study correlation and standard deviations, and another table showing the group-specific parameters, such as the sensitivity and specificity at each level (i.e. group) of the categorical covariate. Furthermore, there will also be a table which displays the pairwise differences and ratios between the pooled sensitivity and specificity estimates (see figure \ref{figure_perfect_gs_meta_regression_parameter_estimates_pairwise_diffs_segment}). 

The 'accuracy vs covariate' subtab contains a plot which displays the summary sensitivity and specificity posterior medians and 95\% credible intervals against the selected covariate. For categorical meta-regression, there will be a posterior median and 95\% credible interval for each category of the covariate, whereas for continuous meta-regression there is a smooth line corresponding to the 95\% posterior median and 95\% credible interval bands as the covariate spans its observed range. 

\begin{figure}[H]
    \centering
    \includegraphics[width=15cm]{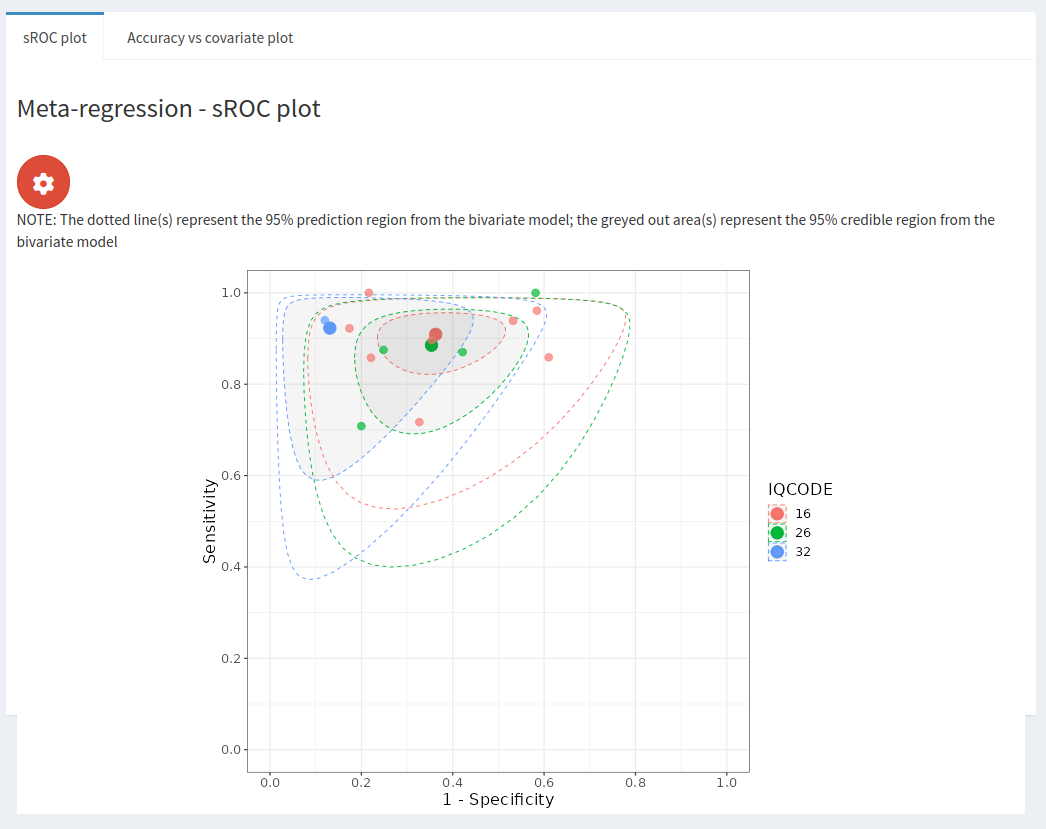}
    \caption{Screenshot of 'Perfect gold standard' tab, 'sROC plot' subtab within 'Meta-regression' subtab}
\label{figure_perfect_gs_meta_analysis_meta_regression_IQCODE_sROC_plot}
\end{figure}

We conducted a categorical meta-regression using the type of IQCODE test (either the 16, 26 or 32-item version) used as the covariate. The results for the 16-item and 26-item groups were very similar (see figure \ref{figure_perfect_gs_meta_regression_parameter_estimates_segment}) - for the 16-item group we obtained sensitivity and specificity estimates of 0.91 (95\% CrI = (0.82, 0.96)) and 0.64 (95\% CrI = (0.50, 0.77)). For the 26-item group we obtained sensitivity and specificity estimates of 0.89 (95\% CrI = (0.72, 0.96)) and 0.65 (95\% CrI = (0.45, 0.82)). For the 32-item group, we obtained a similar sensitivity - 0.92 (95\% CrI = (0.58, 0.99), but for the specificity we obtained a very different result - 0.87 (95\% CrI = (0.58, 0.97)) - however, this was only based on 1 study. Looking at the pairwise differences (see figure \ref{figure_perfect_gs_meta_regression_parameter_estimates_pairwise_diffs_segment}), we can see that the 95\% credible intervals contain 0 for all of the sensitivities and specificities - indicating that none of the differences are significant - even for the comparison to the 32-item group, despite the posterior medians being relatively large. Similarly, the pairwise ratio's all contain 1, implying that none of them are significant. An sROC plot showing the results is shown in figure \ref{figure_perfect_gs_meta_analysis_meta_regression_IQCODE_sROC_plot}.

\subsubsection{Subgroup analysis}
\label{section_perfect_gs_subgroup_analysis}
Our app also allows users to run subgroup analyses for categorical covariate data. This will run a separate bivariate meta-analysis for each subgroup, obviating the need for users to partition their data and run the analysis multiple times. Such analyses differ from including the subgrouping variable as a categorical covariate and using the regression facility outlined above in section \ref{section_perfect_gs_meta_regression}, because here separate random effect variances are calculated for each group, whereas they are assumed to be the same and estimated jointly in the regression. The 'subgroup analysis' tab contains the same subtabs as the 'meta-regression' tab, and the subtabs will look mostly the same as when running a categorical meta-regression. The key difference is that in the 'parameter estimates' subtab, there is just one table showing the parameters for each subgroup, since there are no parameters shared between the subgroups. 

\begin{figure}[H]
    \centering
    \includegraphics[width=15cm]{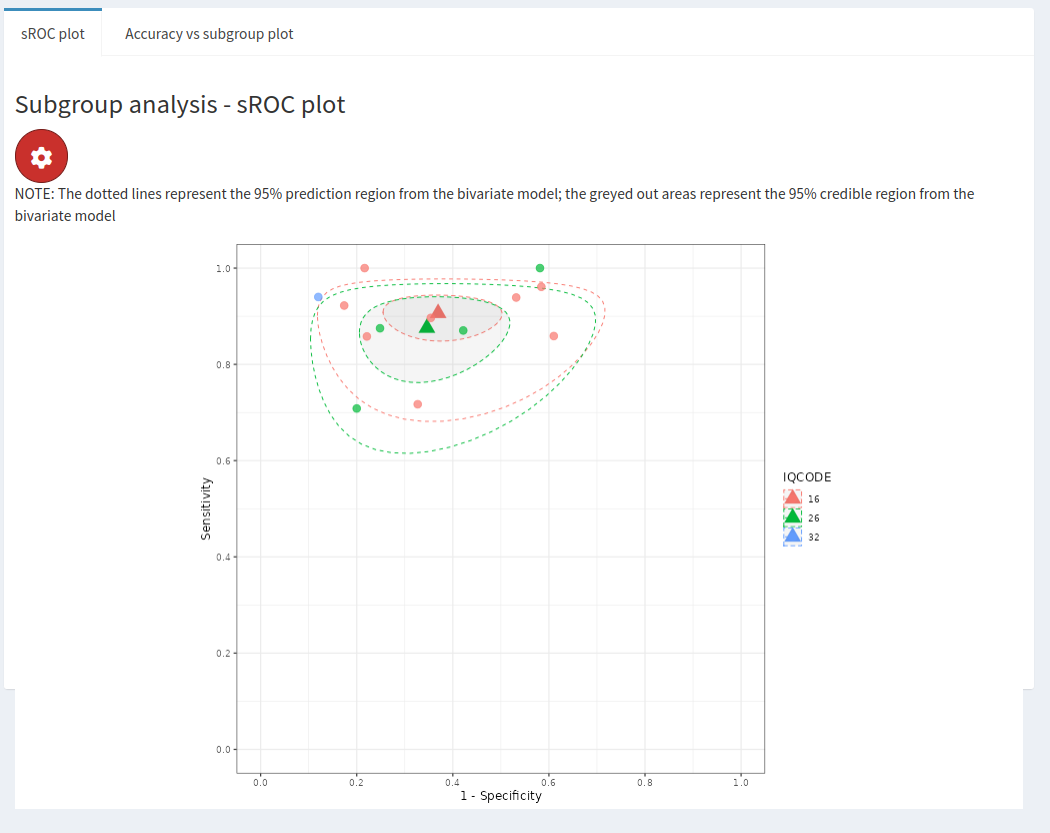}
    \caption{Screenshot of 'Perfect gold standard' tab, 'sROC Plot' subtab within 'subgroup analysis' subtab. Plot corresponds to subgroup analysis for IQCODE test type. }
\label{figure_subgroup_analysis_sroc_plot_iqcode}
\end{figure}

We conducted a subgroup analysis for the type of IQCODE test used - either the 16, 26 or 32-item version. Only one study used the 32-item version, so no analysis could be conducted for this subgroup. Four studies used the 26-item version and eight studies used the 16-item version. The results for these two subgroups were very similar. More specifically, for the 26-item subgroup, we obtained sensitivity and specificity estimates of 0.88 (95\% CrI = (0.77, 0.94)) and 0.65 (95\% CrI = (0.49, 0.79)), and for the 16-item subgroup we obtained sensitivity and specificity estimates of 0.91 (95\% CrI = (0.86, 0.95)) and 0.63 (95\% CrI = (0.51, 0.74)). These results can be compared to the regression demonstration in section \ref{section_perfect_gs_meta_regression}, which made the stronger assumption that the between-study heterogeneity levels are the same across groups. An sROC plot showing the results of the subgroup analysis is shown in figure \ref{figure_subgroup_analysis_sroc_plot_iqcode}.

\subsection{Imperfect gold standard}
\label{section_imperfect_gold_standard}
In addition to meta-analysis of test accuracy which assumes a perfect gold standard using the bivariate model discussed in section \ref{section_perfect_gold_standard}, our app also allows users to run meta-analysis of test accuracy without assuming a perfect gold standard using TLCMs \supercite{Chu2009, Menten2013} within the "Imperfect gold standard" tab. This tab has the following subtabs: 'model set up \& priors', 'Run model', 'study-level outcomes', 'parameter estimates', 'model diagnostics', and 'sROC plot'. 

\begin{figure}[H]
    \centering
    \includegraphics[width=15cm]{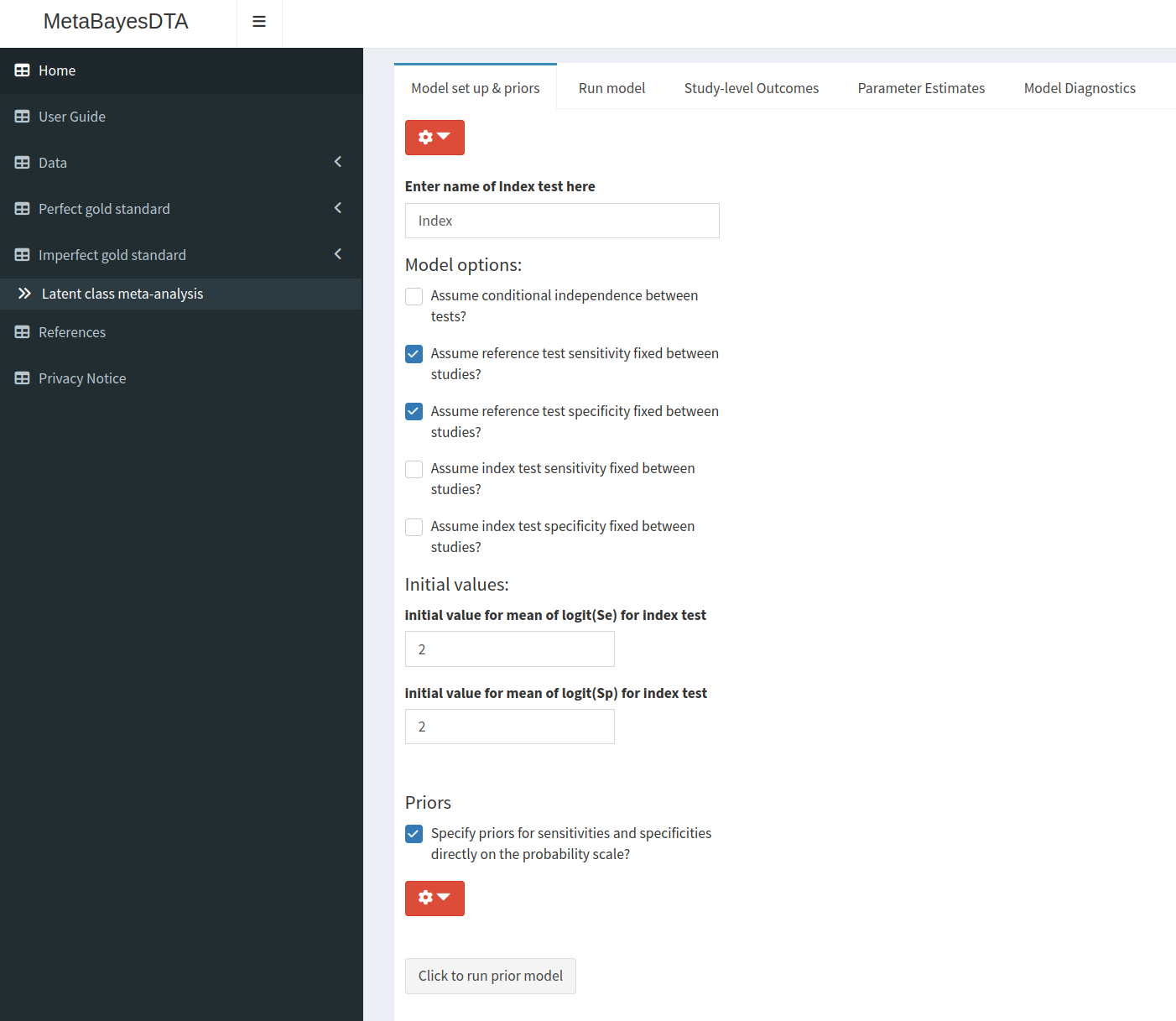}
    \caption{Screenshot of 'Imperfect gold standard' tab, 'Model set up \& priors' subtab within 'Latent class meta-analysis' subtab}
\label{figure_imperfect_gs_model_set_up_and_priors_segment}
\end{figure}

The 'Model set up \& priors' subtab for the TLCM has more options than that of the bivariate model (see figure \ref{figure_imperfect_gs_model_set_up_and_priors_segment}). This is because, in contrast to the bivariate model, which only estimates accuracy for the index test, the TLCM model estimates accuracy for both the index and the reference test(s), as well as the disease prevalence in each study. Users can choose various modelling options - more specifically, they can choose whether the reference and index test sensitivities and specificities are fixed between studies (i.e. "fixed effects"), or whether they can vary between studies (i.e. "random effects"). They can also choose whether to assume \textit{ conditional independence} between tests. In practice, the conditional independence assumption is typically not a reasonable assumption to make, since it assumes that the test results are uncorrelated within the diseased and non-diseased groups \supercite{Vacek1985}. However, sometimes it is not possible to run a model which does not assume conditional independence because it might be \textit{nonidentifiable} \supercite{LCM_models_identifiability}; that is, there might be two (or more) sets of parameter values that fit the data equally well. For instance, the model may estimate the sensitivity of a test to be equal to \textit{both} 0.20 and 0.80. This is more likely to occur when the the number of parameters being estimated from our model is greater than what is possible for the given dataset (although it can also occur when it is possible to estimate all parameters). One way to lower the chance of this happening is to introduce more informative prior information - for instance, information about the accuracy of the reference test(s) is often known and can be obtained by searching the relevant literature and by consulting clinicians. 

In addition to the Stan sampler diagnostics, R-hat statistics, posterior density and trace plots, the 'Model diagnostics' subtab has a plot which allows users to assess the fit of the model - the correlation residual plot \supercite{Qu1996}. 

We conducted an analysis using TLCM models which do not assume a perfect gold standard. The studies included a variety of reference standards - four studies used the Diagnostic and Statistical Manual of Mental Disorders version III, revised (DSM-III-R); seven studies used version IV (DSM-IV); one study used the National Institute of Neurological and Communicative Diseases and Stroke/Alzheimer's Disease and Related Disorders Association (NINCDS-ADRDA) criteria; and one study used a combination of the DSM-III-R and the ICD-10 criteria. Rather than assuming all the reference tests have the same accuracy (as is commonly done in practice), our application allows us to model the differences between the various reference tests using meta-regression. To incorporate prior knowledge into the model, we used information from an umbrella review \supercite{Gaugler2013} (i.e., a review of systematic reviews and meta-analyses). This umbrella review found that the accuracy for clinical dementia diagnostic criteria had a sensitivity range of 0.53-0.93 and a specificity range of 0.55-0.99. For the sensitivities and specificities of all of the reference tests, we used priors corresponding to a 95\% prior interval of $(0.43, 0.96)$.

\subsubsection{Analysis assuming conditional independence}
\label{motivating_example_imperfect_gs_lcm_conditional_independence}
\begin{figure}[H]
    \centering
    \includegraphics[width=12cm]{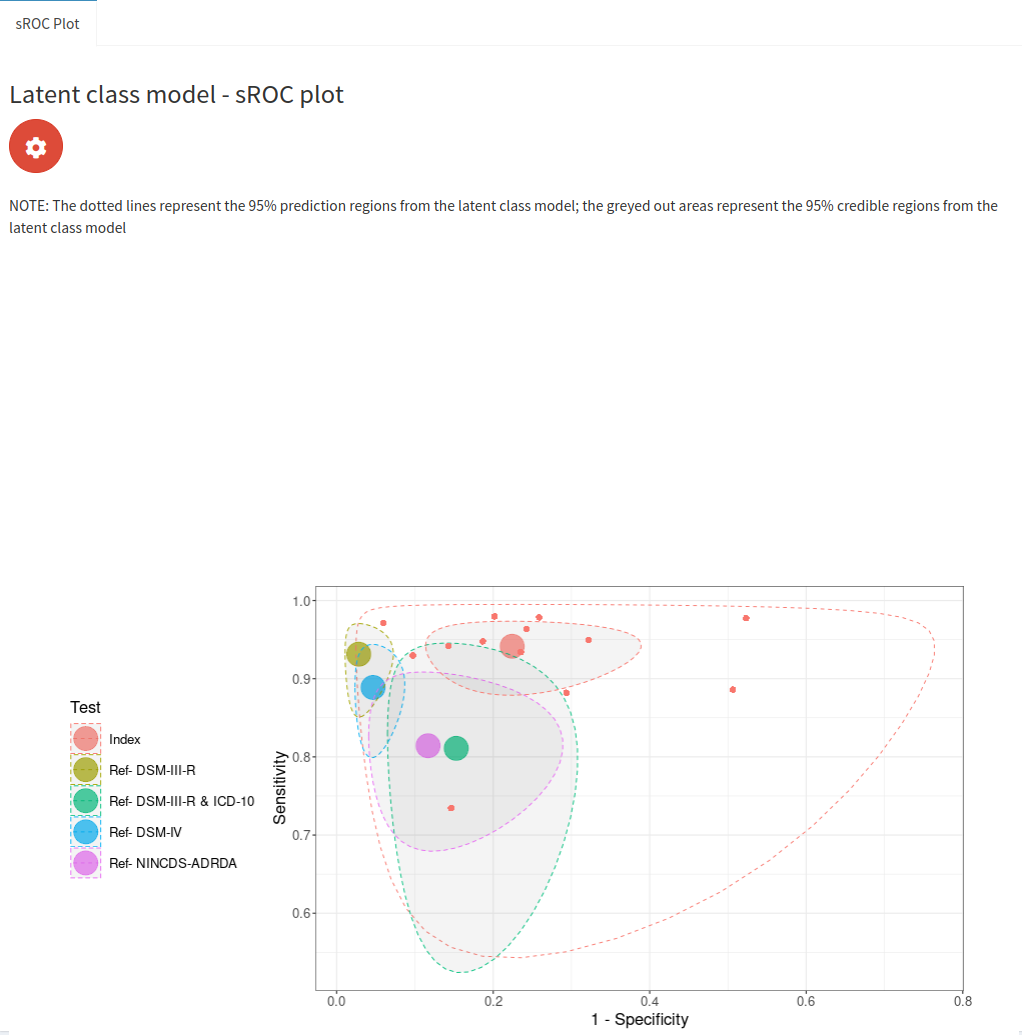}
    \caption{sROC plot for LCM analysis, assuming conditional independence between the IQCODE and reference tests, fixed effects for the reference tests, and random effects for the IQCODE. The red dots correspond to the study-specific estimates for the IQCODE}
\label{imperfect_gs_roc_plot_CI_fixed_refs_random_index}
\end{figure}

\begin{figure}[H]
    \includegraphics[width=15cm]{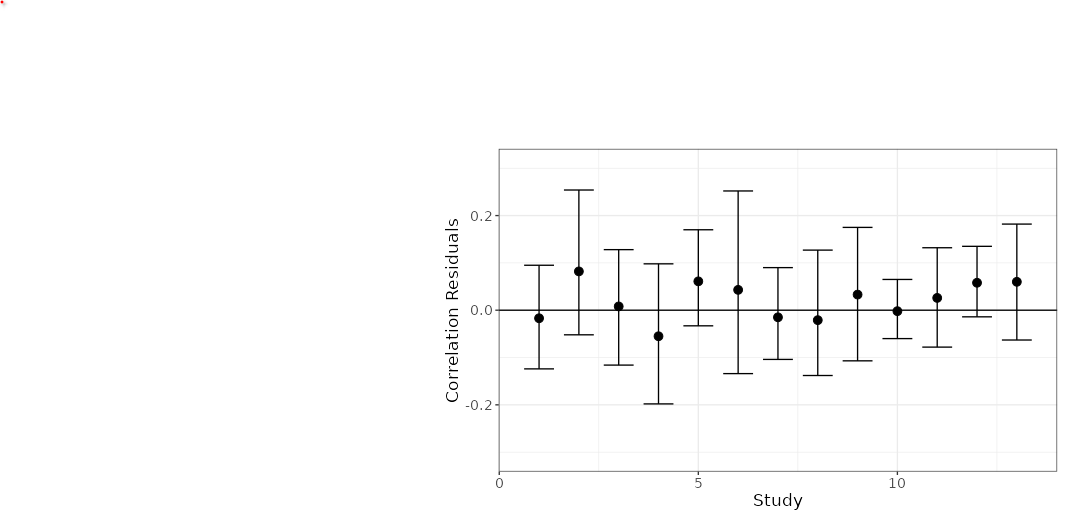}
    \caption{Correlation residual plot for LCM analysis, assuming conditional independence between the IQCODE and reference tests, fixed effects for the reference tests, and random effects for the IQCODE.}
\label{lcm_analysis_CI_fixed_refs_random_index_correlation_residual_plot}
\end{figure}

We first analysed the data using a model which assumed conditional independence between the index test (IQCODE) and the reference tests. We assumed that the reference tests were fixed between studies and assumed random effects for the IQCODE. For the IQCODE, we obtained sensitivity and specificity estimates of 0.94 (95\% CrI = (0.89, 0.98)) and 0.77 (95\% CrI = (0.62, 0.89)). The IQCODE was estimated to have a higher sensitivity but lower specificity than all of the reference tests. These results suggest that the analysis assuming a perfect gold standard (see section \ref{section_perfect_gold_standard}) underestimates the sensitivity of the IQCODE by around 3\% and underestimates the specificity by around 11\%. An sROC plot of the results is shown in figure \ref{imperfect_gs_roc_plot_CI_fixed_refs_random_index}. Attempts to run a model assuming conditional independence between tests with random effects for the reference tests and the index test resulted in unsatisfactory posterior distributions plots (see supplementary material figure \ref{appendix_fig_2}). More specifically, some of the posterior distribution plots for the accuracy parameters were bimodal - that is, they had two means, which means they would estimate the accuracy as being two different values, indicating that the model is non-identifiable. The correlation residual plot (see figure \ref{lcm_analysis_CI_fixed_refs_random_index_correlation_residual_plot}) suggests the conditional independence model provides a satisfactory fit to the data, since all of the 95\% CrI's cross the zero line. 

\subsubsection{Analysis assuming conditional dependence}
\label{motivating_example_imperfect_gs_lcm_conditional_dependence}
\begin{figure}[H]
    \centering
    \includegraphics[width=10cm]{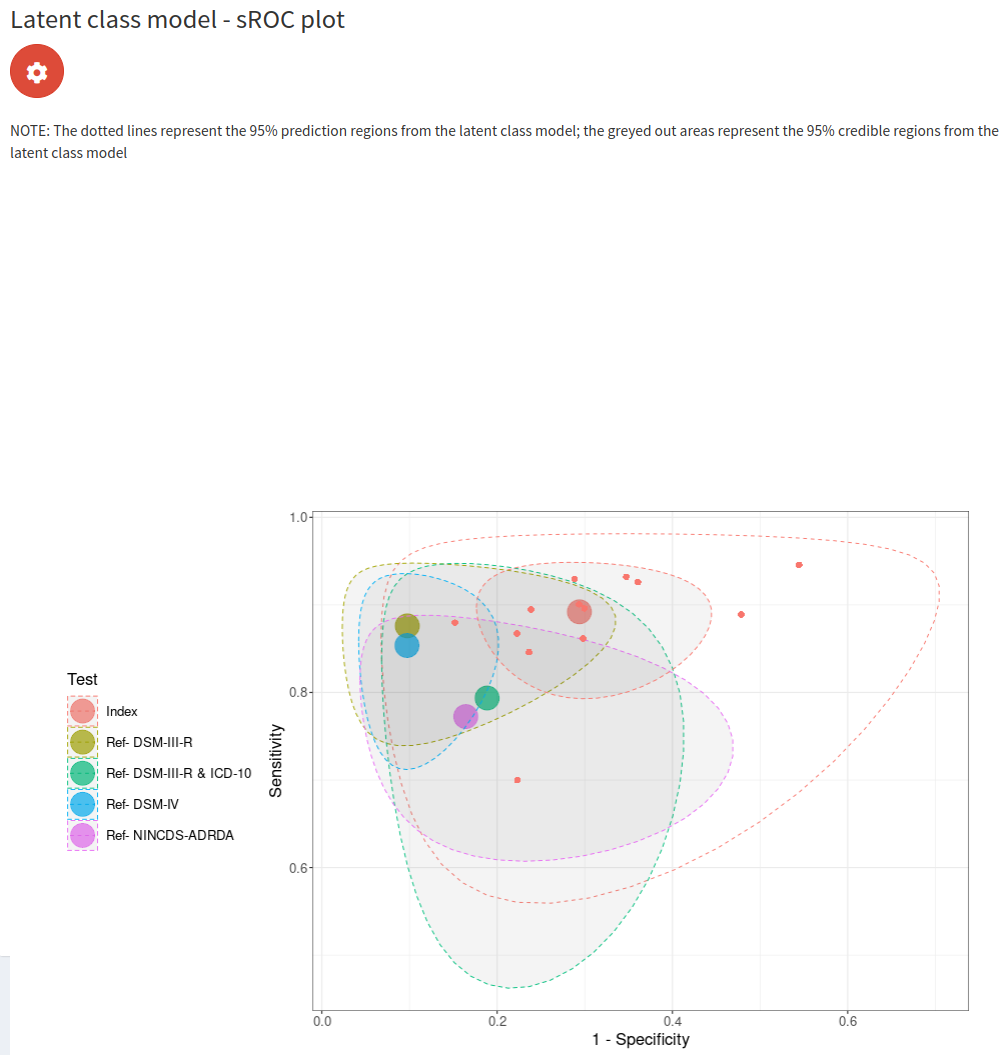}
    \caption{sROC plot for LCM analysis, assuming conditional dependence between the IQCODE and reference tests, fixed effects for the reference tests, and random effects for the IQCODE. The red dots correspond to the study-specific estimates for the IQCODE}
\label{imperfect_gs_roc_plot_CD_fixed_refs_random_index}
\end{figure}

\begin{figure}[H]
    \centering
    \includegraphics[width=10cm]{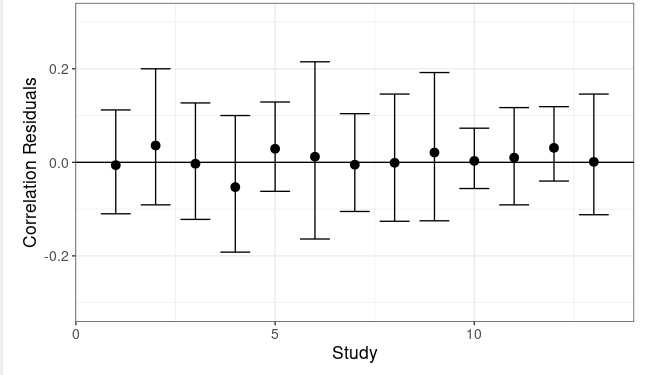}
    \caption{Correlation residual plot for LCM analysis, assuming conditional dependence between the IQCODE and reference tests, fixed effects for the reference tests, and random effects for the IQCODE. The red dots correspond to the study-specific estimates for the IQCODE}
\label{lcm_analysis_CD_fixed_refs_random_index_correlation_residual_plot}
\end{figure}

As mentioned in section \ref{section_imperfect_gold_standard}, despite us obtaining a good correlation residual plot (see figure \ref{lcm_analysis_CI_fixed_refs_random_index_correlation_residual_plot}), the conditional independence assumption is typically not considered to be a reasonable assumption to make in clinical practice. Therefore, we attempted to fit a model without assuming conditional independence between the IQCODE and reference tests. Similarly to the conditional independence case in section \ref{motivating_example_imperfect_gs_lcm_conditional_independence}, there was not enough information to identify all model parameters under the conditional dependence assumption if random effects were assumed for all tests (see supplementary material figure \ref{appendix_fig_3} ); therefore, we made the stronger assumption specifying fixed effects for the reference tests. For the IQCODE, we obtained sensitivity and specificity estimates of 0.89 (95\% CrI = (0.82, 0.95)) and 0.71 (95\% CrI = (0.58, 0.84)). Both of these estimates are lower than the model assuming conditional independence (see section \ref{motivating_example_imperfect_gs_lcm_conditional_independence}), and suggest that the analysis assuming a perfect gold standard slightly overestimated the sensitivity by around 2\% and underestimated the specificity by around 5\%. An sROC plot of the results is shown in figure \ref{imperfect_gs_roc_plot_CD_fixed_refs_random_index}. Furthermore, although the model with conditional independence provided a satisfactory fit (see figure \ref{lcm_analysis_CI_fixed_refs_random_index_correlation_residual_plot}), the conditional dependence model clearly provides a better fit (see figure \ref{lcm_analysis_CD_fixed_refs_random_index_correlation_residual_plot}) - since it moves the summary estimates of the residual correlations closer to 0. 

\section{Discussion}
In this paper, we presented MetaBayesDTA, an extensively expanded web-based R Shiny \supercite{shiny_r_package} application based on MetaDTA \supercite{metadta_Freeman2019-mk}. The application enables users to conduct Bayesian meta-analysis of diagnostic test accuracy studies, both assuming a perfect reference test or modelling an imperfect reference test, without users having to install any software or have any knowledge of R \supercite{R_software_ref} or Stan \supercite{stan} programming. 

The application uses the bivariate model \supercite{Reitsma2005} to conduct analysis assuming a perfect reference test, and users can also conduct univariate meta-regression and subgroup analysis. It uses TLCMs \supercite{Chu2009, Menten2013} to conduct analyses without assuming a perfect gold standard, allowing the user to run models assuming conditional independence or dependence, options for whether to model the reference and index test sensitivities and specificities as fixed or random effects, and can model multiple reference tests using a meta-regression covariate for the type of reference test. The application allows users to input their own prior distributions, which is particularly useful for the TLCM models since information about the accuracy of the reference test(s) is often known. Similarly to MetaDTA \supercite{metadta_Freeman2019-mk}, the tables and figures can be downloaded, and the graphs are highly customizable. Furthermore, risk of bias and quality assessment results from the QUADAS-2 \supercite{quadas_2_tool} tool can be incorporated into the sROC plot; integrating risk of bias into the main analysis decreases the tendency to think of risk of bias as an afterthought. Sensitivity analysis allowing users to remove selected studies can also be carried out easily for all models. 

As we discussed in section \ref{section_why_app_needed} (see table \ref{table_app_comparison}), our app offers improvements over both BayesDTA (\url{https://bayesdta.shinyapps.io/meta-analysis}) and MetaDTA \supercite{metadta_Freeman2019-mk}. Namely, for the bivariate model, unlike both BayesDTA and MetaDTA, our app allows subgroup analysis and univariate meta-regression (either categorical or continuous covariate) to be carried out, which also allows users to easily conduct comparative test accuracy meta-analysis to compare two or more tests to one another. Furthermore, unlike BayesDTA, for the TLCM, our app can assess model fit using the correlation residual plot \supercite{Qu1996}, and it can model multiple reference tests, using a categorical covariate for the type of reference test. This is important since studies included in meta-analysis of test accuracy often use different reference tests, and the accuracy can vary greatly between them. Even though it is more complicated than MetaDTA, as it can run 5 different models rather than one and the graphs have more customization options, it has a cleaner layout and many of the menus are hidden unless the user clicks on them to display more options, thanks to the shinydashboard \supercite{shinydashboard_r_package} and shinyWidgets\supercite{shinywidgets_r_package} R packages. 

Our web application has some limitations which give way to future developments. For example relating to meta-analysis of test accuracy without assuming a perfect gold standard (TLCM), whilst users can model the data without assuming conditional independence between tests, it does not offer functionality to impose restrictions on the correlation structure. Therefore, a potential improvement would be to allow users to impose these restrictions, such as assuming the same correlation in the diseased and non-diseased groups, and/or forcing the correlations between the tests to be positive. Furthermore, for the TLCM, although our application allows users to assess model fit using the correlation residual plot \cite{Qu1996}, other measures of model fit - such as the deviance and observed verses expected 2x2 tables of test results - are not yet available. Another limitation of the TLCM is that it can only model different reference tests using categorical meta-regression and therefore assumes that all of the reference tests have the same between-study variances. Although this is often an advantage compared to conducting a subgroup analysis for each reference test, sometimes it might make sense to run a more complex model which assumes separate between-study variances for some reference tests and assumes fixed effects for reference tests only observed in a few (e.g., 5) studies, therefore adding this functionality is a potential update. 

For the bivariate model, a potential update for both subgroup analysis and categorical meta-regression would be allow users to specify different priors for each of the groups. Furthermore, for meta-regression, although our application allows users to see the pairwise differences and ratio's between the different categories of a categorical covariate (making it possible to use for comparative test accuracy of multiple tests), it only shows these for the meta-regression which assumes the variances are the same between all tests. However, in some instances it might make sense for the variances for some (or all - which would be equivalent to conducting a subgroup analysis) of the tests to be different, so a future update to improve the application would be to also display the pairwise differences and ratios for the subgroup analysis, and allowing users to assume independent variances for some tests but shared variances across other tests. 

Another limitation is that our application only allows subgroup analysis and meta-regression (besides for modelling different reference tests) to be conducted using the bivariate model, which assumes a perfect gold standard. A potential improvement would be to allow users to run subgroup analyses and meta-regression for the TLCM. Furthermore, the application requires users to have some knowledge about checking Bayesian model diagnostics to check that the models have been fitted OK - although the application does contain some information (in the "model diagnostics" tabs) which explains how to interpret some of the model diagnostics, and also directs users to online resources which explain how to interpret the model diagnostics so users do not have to find this information themselves. 

It is important to note that this app is a beta version, so it is expected that there may be some bugs. Therefore, we welcome any user feedback (by emailing the first author of this paper) to enable MetaBayesDTA to develop into a user-friendly and widely used diagnostic test accuracy meta-analysis web application, as MetaDTA \supercite{metadta_Freeman2019-mk} has become.

In general, one could argue that easy-to-use apps could lead to the over-application of complex methods even when they are not appropriate. This is because web applications - such as the one presented in this paper - will allow less experienced researchers to be able to conduct complex analyses which would otherwise be inaccessible to them, lowering the amount of knowledge needed to perform the analysis, and therefore increasing the chance of invalid results being published. Therefore, we recommend that there is a statistician (with knowledge of how to check Bayesian model diagnostics) in the review team. However, one could also argue that the widespread usability of apps could stimulate the uptake of more appropriate methods, which means that better methods will become standard practice more quickly. This could have important impacts for clinical practice; for instance, the fact that our app allows one to easily conduct a meta-analysis of test accuracy without assuming a gold standard without assuming the same reference test is used across all studies opens up many new datasets to synthesis, since many studies are conducted using different imperfect reference tests. 

\section{Conclusions}
In this paper, we presented MetaBayesDTA (\url{https://crsu.shinyapps.io/MetaBayesDTA/}), a user-friendly, interactive web application which allows users to conduct Bayesian meta-analysis of test accuracy, with or without a gold standard. The application uses methods which were previously only available by using statistical programming languages, such as R \supercite{R_software_ref}. 

This application could have a wide-ranging impact across academia, guideline writers, policy makers, and industry. For example, when there is not a perfect reference test available, the estimates of test accuracy can change quite notably when relaxing the perfect reference test assumption, leading to potentially different conclusions being drawn about the accuracy of a test which could ultimately lead to changes in which tests are used in clinical practice. Furthermore, the ability of the app to easily conduct comparative test accuracy meta-analysis means that clinicians will more easily be able to tell which tests are better. 


%% file: Files/Highlights_acknl_data.tex
\section*{Highlights}
\subsection*{What is already known?}
\begin{itemize}
  \item Conducting meta-analysis of test accuracy requires relatively high level of specialised knowledge. 
  \item Recently, web applications such as MetaDTA (\url{https://crsu.shinyapps.io/dta_ma/}) and BayesDTA (\url{https://bayesdta.shinyapps.io/meta-analysis}) have been released which allow users to conduct such analyses without having to have as much training in statistical programming. 
  \item These applications have several limitations. 
\end{itemize}

\subsection*{What is new?}
\begin{itemize}
\item The application is based on MetaDTA, and addresses several key limitations of other applications, for example:
\item For the meta-analysis of test accuracy assuming a gold standard using the bivariate model, one can conduct subgroup analysis, univariate meta-regression with either a categorical or continuous covariate, and comparative test accuracy evaluation. 
\item For the model which does not assume a perfect gold standard, the application can account for the fact that different studies in a meta-analysis use different reference tests. 
\end{itemize}

\subsection*{Potential impact}
\begin{itemize}
\item Thanks to its user-friendly and interactive layout, we believe that MetaBayesDTA will be used by applied researchers from a variety of fields.
\item This is because the ability to conduct a meta-analysis of test accuracy is important not only for clinical research but also for guideline writers, policy makers, and industry. 
\end{itemize}

\section*{Acknowledgements}
The authors would like to thank Olivia Carter for carefully proofreading the manuscript. 

Funding: The work was carried out whilst EC was funded by a National Institute for Health Research (NIHR) Complex Reviews Support Unit (project number 14/178/29) and by an NIHR doctoral research fellowship (project number NIHR302333). The views and opinions expressed herein are those of the authors and do not necessarily reflect those of the NIHR, NHS or the Department of Health. The NIHR had no role in the design of the study and collection, analysis, and interpretation of data and in writing the manuscript. This project is funded by the NIHR Applied Research Collaboration East Midlands (ARC EM). The views expressed are those of the authors and not necessarily those of the NIHR or the Department of Health and Social Care. 

\section*{Data availability statement}
The web application is available at: 
\href{https://crsu.shinyapps.io/MetaBayesDTA/}{https://crsu.shinyapps.io/MetaBayesDTA/}.  
The data, R and Stan code for the web application is available at: 
\href{https://github.com/CRSU-Apps/MetaBayesDTA}{https://github.com/CRSU-Apps/MetaBayesDTA}.

%% file: Files/Appendix.tex
\label{appendix}
\begin{figure}[H]
    \centering
    \includegraphics[width=15cm]{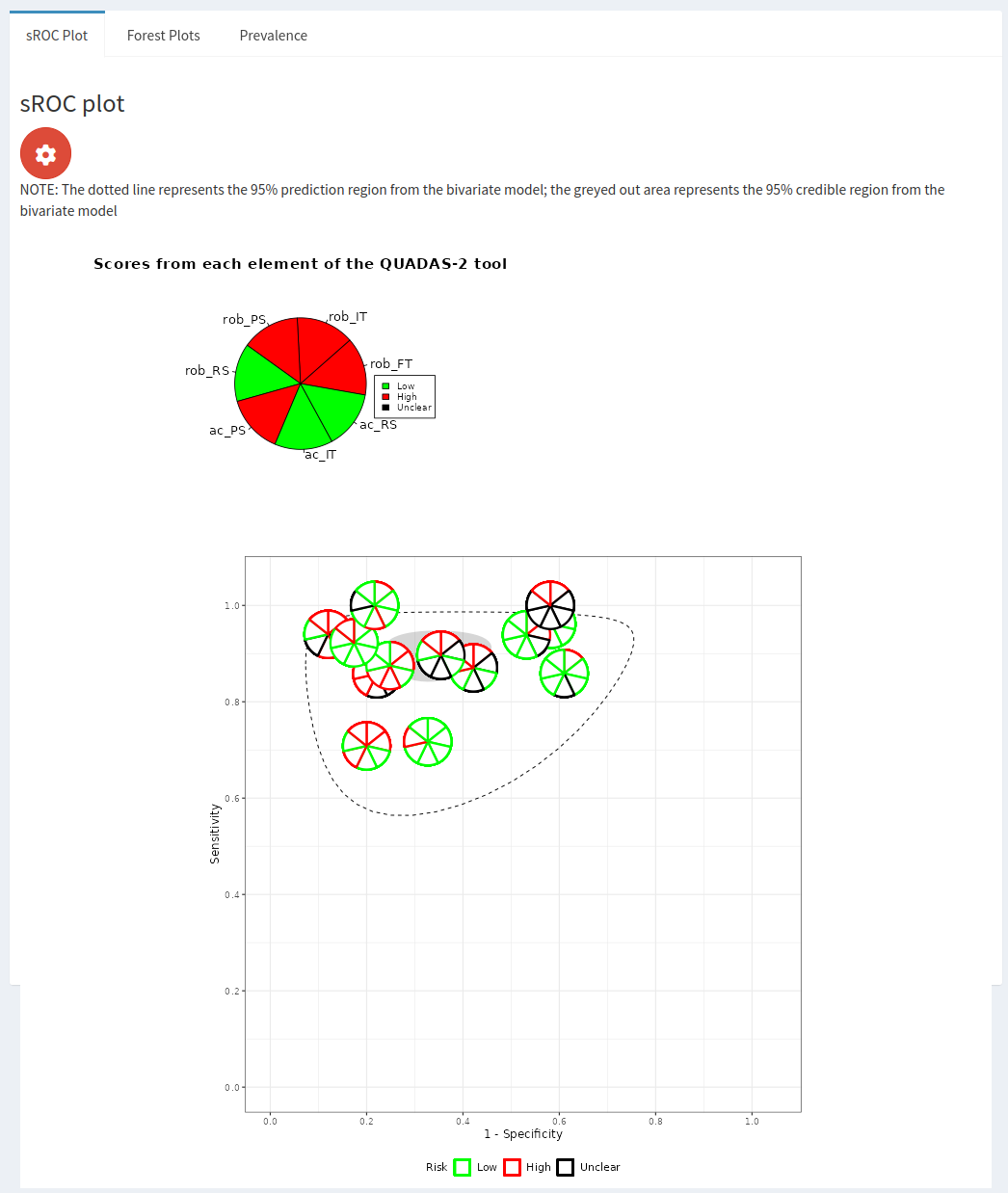}
    \caption{Posterior density plots for model assuming random effects for index test (MMSE) and reference tests and conditional independence between tests.}
    \label{appendix_fig_1}
\end{figure}

\begin{figure}[H]
    \centering
    \includegraphics[width=15cm]{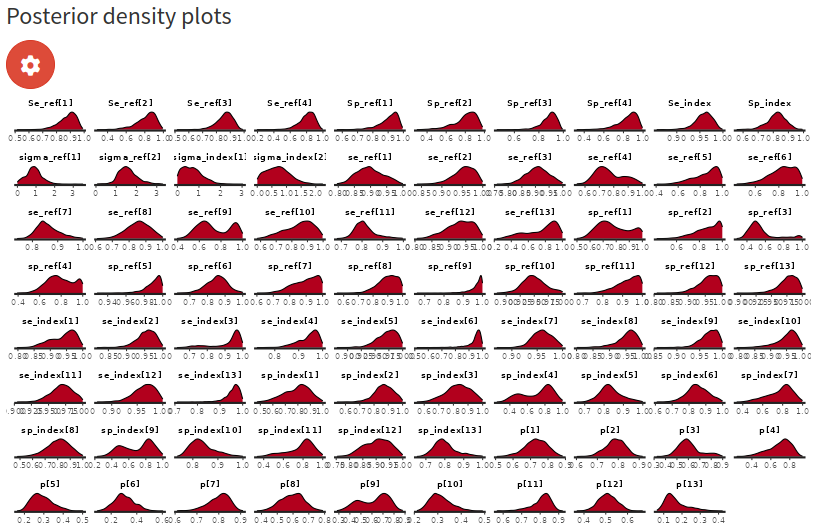}
    \caption{Posterior density plots for model assuming random effects for index test (MMSE) and reference tests and conditional independence between tests.}
    \label{appendix_fig_2}
\end{figure}

\begin{figure}[H]
    \centering
    \includegraphics[width=15cm]{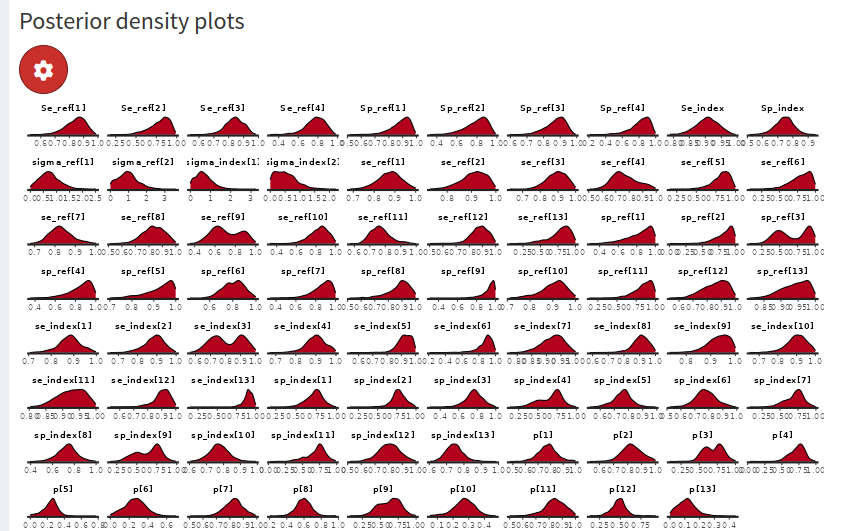}
    \caption{Posterior density plots for model assuming random effects for index test (MMSE) and reference tests and conditional dependence between tests.}
    \label{appendix_fig_3}
\end{figure}